\documentclass[twocolumn,conference]{IEEEtran}
\usepackage[T1]{fontenc}
\usepackage[latin9]{inputenc}
\usepackage{color}
\usepackage{prettyref}
\usepackage{calc}
\usepackage{amsmath}
\usepackage{amssymb}
\usepackage{graphicx}

\makeatletter
\usepackage{liulyx}
\usepackage{acronym}
\AtBeginDocument{\acrodef{ASP}{antenna separation product}
\acrodef{AWGN}{additive white Gaussian noise}
\acrodef{BEP}{bit error probability}
\acrodef{BER}{bit error rate}
\acrodef{BF-MIMO}[BF\mbox{-}MIMO]{beamforming MIMO}
\acrodef{BF}{beamforming}
\acrodef{bpcu}{bits per channel use}
\acrodef{CP}{cyclic prefix}
\acrodef{CSI}{channel state information}
\acrodef{CSIR}{channel state information at RX}
\acrodef{SSK}{space shift keying}
\acrodef{CSIT}{channel state information at TX}
\acrodef{DCMC}{discrete\mbox{-}input continuous\mbox{-}output memoryless channel}
\acrodef{DFT}{discrete Fourier transform}
\acrodef{DL-TR-GSM}{dual-layered transmit-receive \acl{GSM}}
\acrodef{DLT}{dual-layered transmission}
\acrodef{EGC}{equal gain combining}
\acrodef{EM}{electromagnetic}
\acrodef{FSPL}{free space path loss}
\acrodef{FFT}{fast Fourier transform}
\acrodef{FDE}{frequency domain equalization}
\acrodef{GRSM}{generalized \acl{RSM}}
\acrodef{GSM}{generalized \acl{SM}}
\acrodef{IFFT}{invserse fast Fourier transform}
\acrodef{ICI}{inter-channel interference}
\acrodef{iid}[i.i.d.]{independent and identically distributed}
\acrodef{IQ}{in\mbox{-}phase and quadrature}
\acrodef{ISI}{intersymbol interference}
\acrodef{ISI-free}[ISI\mbox{-}free]{intersymbol interference free}
\acrodef{LIS}{large intelligent surface}
\acrodef{LOS}{line\mbox{-}of\mbox{-}sight}
\acrodef{mmWave}{millimeter-wave}
\acrodef{MIMO}{multiple\mbox{-}input multiple\mbox{-}output}
\acrodef{MISO}{multiple\mbox{-}input single\mbox{-}output}
\acrodef{ML}{maximum likelihood}
\acrodef{MRC}{maximal ratio combining}
\acrodef{MMSE}{minimum mean square error}
\acrodef{MU-TR-GSM}{multiuser transmit-receive  \acl{GSM} }
\acrodef{NCSIT}{no channel state information at TX}
\acrodef{NLOS}{non\mbox{-}\acs{LOS}} 
\acrodef{OFDM}{orthogonal frequency division multiplexing}
\acrodef{PA}{power amplifier}
\acrodef{PAE}{power added efficiency}
\acrodef{PAPR}{peak\mbox{-}to\mbox{-}average power ratio}
\acrodef{PDF}{probability density function}
\acrodef{PEP}{pairwise error probability}
\acrodefplural{PEP}{pairwise error probabilities}
\acrodef{PMP}{probability mass function}
\acrodef{PSM}{precoding-aided spatial modulation}
\acrodef{QSM}{quadrature spatial modulation}
\acrodef{RC}{reorganization computation}
\acrodef{RIS}{reconfigurable intelligent surface}
\acrodef{RSM}{receive spatial modulation}
\acrodef{RX}{receiver}
\acrodef{SEP}{symbol error probability}
\acrodef{SER}{symbol error rate}
\acrodef{SM}{spatial modulation}
\acrodef{SMX-MIMO}[SMX\mbox{-}MIMO]{spatial multiplexing MIMO}
\acrodef{SMX}{spatial multiplexing}
\acrodef{SNR}{signal-to-noise ratio}
\acrodef{SC}{single carrier}
\acrodef{SISO}{single-input single-output}
\acrodef{SVD}{singular value decomposition}
\acrodef{SPST}{single pole single-throw}
\acrodef{TDE}{time domain equalization}
\acrodef{TX}{transmitter}
\acrodef{ULA}{uniform linear array}
\acrodef{VGA}{variable gain amplifier}
\acrodef{ZF}{zero-forcing}
\acrodef{ZMCG}{zero-mean complex Gaussian}

}
 
\@ifundefined{showcaptionsetup}{}{%
 \PassOptionsToPackage{caption=false}{subfig}}
\usepackage{subfig}
\makeatother

\begin{document}
\title{Channel Capacity Optimization Using Reconfigurable Intelligent Surfaces
in Indoor mmWave Environments}
\author{\IEEEauthorblockN{Nemanja~Stefan~Perovi\'c\IEEEauthorrefmark{1}, Marco~Di~Renzo\IEEEauthorrefmark{2},
and Mark~F.~Flanagan\IEEEauthorrefmark{1}}\IEEEauthorblockA{\IEEEauthorrefmark{1}School of Electrical and Electronic Engineering,
University College Dublin\\
Belfield, Dublin 4, Ireland\\
Email: nemanja.stefan.perovic@ucd.ie, mark.flanagan@ieee.org}\IEEEauthorblockA{\IEEEauthorrefmark{2}Laboratoire des Signaux et Systèmes, CNRS, CentraleSupélec,
Université Paris-Saclay\\
3 rue Joliot Curie, Plateau du Moulon, 91192, Gif-sur-Yvette, France\\
Email: marco.direnzo@centralesupelec.fr}}
\maketitle
\begin{abstract}
Indoor \ac{mmWave} environment channels are typically sparsely-scattered
and dominated by a strong \ac{LOS} path. \textcolor{black}{Therefore,
communication over such channels is in general extremely difficult
when the \ac{LOS} path is not present. However, the recent introduction
of \acp{RIS}, which have the potential to influence the propagation
environment in a controlled manner, has the potential to change the
previous paradigm. Motivated by this, we study the channel capacity
optimization utilizing \acp{RIS} in indoor }\ac{mmWave}\textcolor{black}{{}
environments where no LOS path is present. More precisely, we propose
two optimization schemes that exploit the customizing capabilities
of the \ac{RIS} reflection elements in order to maximize the }channel
capacity. The first optimization scheme exploits only the adjustability
of the RIS reflection elements; for this scheme we derive an approximate
expression which explains the connection between the channel capacity
gains and the system parameters. The second optimization scheme jointly
optimizes the RIS reflection elements \emph{and} the transmit phase
precoder; for this scheme, we propose a low-complexity technique called
\emph{global co-phasing} to determine the phase shift values for use
at the RIS. Simulation results show that the optimization of the RIS
reflection elements produces a significant channel capacity gain,
and that this gain increases with the number of \ac{RIS} elements.\textcolor{blue}{{}
}\acresetall{}
\end{abstract}


\section{Introduction}

\textcolor{black}{\bstctlcite{BSTcontrol}}Over the last years, there
has been a tremendous, almost exponential, increase in the demands
for higher data rates. The main driving forces that constantly enhance
these demands are the increasing number of mobile-connected devices
and the appearance of services that require high data rates (e.g.,
video streaming, online gaming). Consequently, a key feature of novel
wireless communication standards is the migration to higher frequencies,
such as the \ac{mmWave} frequency bands. These bands provide us with
multi-GHz bandwidths and enable transmission data rates of multi-Gbps.

Among the proposed \ac{mmWave} frequency bands, the 60$\,$GHz band
with up to 7$\,$GHz (57-64$\,$GHz) of available bandwidth worldwide
\cite{Daniels2010} has gained significant interest in industry as
well as in academia. Because of the large signal attenuation, caused
by the high \ac{FSPL} and the oxygen absorption, communication in
the 60$\,$GHz band is implementable in practice only for short-range
distances and will most likely be used for indoor environments \cite{Yang2008}.
The main characteristic of indoor mmWave communications is that the
propagation channel exhibits a quasi-optical behavior with most of
the signal energy being received along the \ac{LOS} path\cite{Maltsev2010}.
This is due to high re\textcolor{black}{flection losses and high path
losses caused by the larger path length compared to the \ac{LOS}
path. However, the presence of different obstacles (e.g., furniture,
walls) may easily block the LOS path. In this scenario, if the number
of metallic surfaces (or surfaces with relatively low reflection losses)
is low, it can be extremely difficult to establish a reliable communication
link.}

\textcolor{black}{A possible approach to overcome this issue lies
in the use of }\acp{RIS}. \acp{RIS} transform a generally stochastic
channel into a software-reconfigurable environment that actively participates
in transmitting and processing information \cite{basar2019wireless}.
The key component to realize the RIS function is a software-defined
meta-surface that is reconfigurable in a way to adapt itself to changes
in the wireless environment \cite{ntontin2019reconfigurable,di2019smart}.
RISs consist of a large number of small, low-cost, and passive elements
each of which can reflect the incident signal with an adjustable phase
shift, thereby modifying the radio wave. Optimization of the wavefront
of the reflected signals enables us to shape how the radio waves interact
with the surrounding objects, and thus  control their scattering
and reflection characteristics. For this reason, the focus of this
paper is on the utilization of the RIS elements' phase adjustment
capabilities to optimize the system performance in indoor \ac{mmWave}
communications.

A body of research work has very recently emerged which studies the
design of \ac{RIS}-assisted \ac{MIMO} systems for conventional (i.e.,
non-mmWave) channels. In \cite{wu2019intelligent}, an optimization
scheme to enhance the receive \ac{SNR} of a single-stream \ac{MISO}
system by jointly adjusting the RIS reflection elements and the transmit
precoder was proposed. However, this optimization scheme is constrained
to single-antenna receivers. An asymptotic analysis of \acp{RIS}
in \ac{MIMO} multi-user systems was presented in \cite{nadeem2019large}
and it was shown that \acp{RIS} enable performance gains that are
comparable to the gains of massive \ac{MIMO} systems. The energy-efficiency
of \acp{RIS} in multi-user communication was studied in \cite{huang2019reconfigurable},
and energy-efficient designs for both the transmit power allocation
and the phase shifts of the \ac{RIS} elements were developed. In
\cite{basar2019large}, the authors extended the concept of \ac{RIS}
communications to the realm of index modulation by introducing RIS
\ac{SSK} and RIS \ac{SM} schemes.

In spite of this high research interest, only a s\textcolor{black}{mall
number of papers consider the use of \acp{RIS} in \ac{mmWave} communications.
In \cite{tan2018enabling}, the authors investigated the blockage
problem and used an RIS to optimize the link outage probability for
\ac{SISO} systems in indoor \ac{mmWave} environments. This approach
was also validated by a hardware testbed for SISO communication via
an RIS. However, the work of \cite{tan2018enabling} only considered
the scenario where the transmitter and receiver are each equipped
with a single antenna. Also, the system design in \cite{tan2018enabling}
did not consider the use of phase precoding at the transmitter. The
fundamental limits of utilizing \acp{RIS} in mmWave \ac{MIMO} positioning
systems were investigated in }\cite{he2019large}. The optimization
scheme for a single-stream \ac{MISO} transmission in \ac{mmWave}
communication systems was studied in \cite{wang2019intelligent}.
In that paper, an expression for computing the optimal transmit precoding
vector was derived. Since the hardware implementation of this precoding
vector required the fully-digital hardware architecture which has
a prohibitively high hardware complexity, the authors considered a
mathematical approximation of this precoding vector that is implementable
by the hybrid hardware architecture. Both transmit precoder hardware
architectures require the use of \acp{VGA}, which usually have a
much higher hardware complexity than conventional amplifiers, to adjust
the signal amplitude. Also, the aforementioned approximation is
usually performed using a compressed sensing algorithm, which can
significantly increase the computational complexity. In addition,
the optimization scheme in \cite{wang2019intelligent} is also constrained
to single-antenna receivers.

Against this background, the contributions of this paper can be summarized
as follows:
\begin{itemize}
\item We propose two schemes for channel capacity optimization for single-stream
\ac{MIMO} transmission systems utilizing \acp{RIS} in indoor \ac{mmWave}
environmen\textcolor{black}{ts environments where no LOS path is present.}
Although these schemes do not in general yield the global optimum
solution, they are based purely on signal phase adjustments and hence
require only low-complexity hardware architectures, which makes them
particularly suitable for \ac{mmWave} communications.
\item The first optimization scheme utilizes only the adjustability of the
RIS reflection elements for which we derive the optimal phase shift
values. Also, we derive an approximate expression that simplifies
the channel capacity calculation for this optimization scheme, and
intuitively explains the connection between channel capacity gains
and the system parameters.
\item The second optimization scheme utilizes jointly the adjustability
of the RIS reflection elements and the transmit phase precoder to
optimize the channel capacity. For this scheme, we develop a low-complexity
technique called \emph{global co-phasing}, to determine the appropriate
phase shift values of the RIS reflection elements.
\item We show through channel capacity simulations that the proposed optimization
schemes are\textcolor{black}{{} particularly suitable} for indoor \ac{mmWave}
communication systems with a large number of \ac{RIS} elements.
\end{itemize}
\emph{Notation}: Lowercase bold symbols denote column vectors; uppercase
bold symbols denote matrices; $\left|\cdot\right|$ and $\left\Vert \cdot\right\Vert $
denote absolute value and $\mathrm{L_{2}}$-norm, respectively; $\arg\{\cdot\}$
and $\mathbb{E}\{\cdot\}$ denote the argument of a complex number
and the mean (expected) value of a random variable, respectively;
$(\cdot)^{\mathrm{T}}$ and $(\cdot)^{\mathrm{H}}$ denote transpose
and Hermitian transpose, respectively; $j$ denotes the imaginary
unit; $A(i,k)$ denotes the \emph{k}-th element of the \emph{i}-th
row of matrix $\mathbf{A}$; $\mathbf{b}_{*,i}$ denotes the \emph{i}-th
column of matrix $\mathbf{B}$; $\mathbf{c}_{i,*}$ denotes the \emph{i}-th
row of matrix $\mathbf{C}$; $\mathcal{CN}(\mu,\sigma^{2})$ denotes
a circularly symmetric complex Gaussian random variable of mean $\mu$
and variance $\sigma^{2}$.

\section{System Model}

\begin{figure}[t]
\centering{}\includegraphics[height=4.9cm]{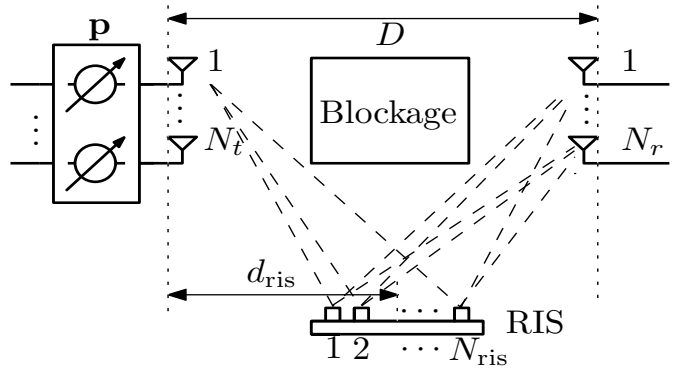}\caption{Illustration of the considered communication system. \label{fig:Block-diag-RIS}}
\end{figure}
We consider a wireless communication system with $N_{t}$ transmit
and $N_{r}$ receive antennas, which is depicted in \prettyref{fig:Block-diag-RIS}.
Both the transmit and receive anten\textcolor{black}{nas are placed
in \acp{ULA} on vertical walls that are parallel to each other. The
distance between these walls is denoted as $D$. The heights of the
transmit and the receive antenna array midpoints are $h_{t}$ and
$h_{r}$ respectively. In simulations (see \prettyref{sec:Simulation-Results}),
$h_{t}$ and $h_{r}$ will take values from predefined uniform ranges
whose mean values are $h_{t,m}$ and $h_{r,m}$ respectively. The
inter-antenna separations of the transmit and the receive antenna
array are $s_{t}$ and $s_{r}$ respectively. Also, one RIS is installed
on the surface perpendicular to the antenna arrays (e.g., floor or
other wall). It consists of $N_{\mathrm{ris}}$ reflection elements
placed uniformly in one dimension and the separation between the adjacent
RIS elements is $s_{\mathrm{ris}}$. T}he distance between the midpoint
of the RIS and the plane containing the transmit array is $d_{\mathrm{ris}}$.
We assume that the RIS elements are ideal and that each of them can
independently influence the phase and the reflection angle of the
impinging wave.

As the focus of this paper is on a single \ac{IQ} stream transmission
in the considered communication system, the signal vector at the receive
antennas is given by 
\begin{equation}
\mathbf{y}=\frac{1}{\sqrt{N_{t}}}\mathbf{Hp}s+\mathbf{n},\label{eq:ss_equ}
\end{equation}
where $\mathbf{H}\in\mathbb{C}^{N_{r}\times N_{t}}$ is the channel
matrix, $s$ is the transmitted \ac{IQ} symbol with the energy $E_{s}=\mathbb{E}\{\left|s\right|^{2}\}$
and $\mathbf{p}=[p_{1}\;\cdots\;p_{N_{t}}]^{\mathrm{T}}\in\mathbb{C}^{N_{t}\times1}$
is the transmit precoding vector which satisfies $\left\Vert \mathbf{p}\right\Vert ^{2}=N_{t}$.
The noise vector $\mathbf{n}\in\mathbb{C}^{N_{r}\times1}$ is distributed
according to $\mathcal{CN}(\mathbf{0},N_{0}\mathbf{I})$.

\textcolor{black}{The $(r,t)$ element of the channel matrix $\mathbf{H}$
that models signal propagation between the }\textcolor{black}{\emph{t}}\textcolor{black}{-th
transmit antenna and the }\textcolor{black}{\emph{r}}\textcolor{black}{-th
receive antenna via all $N_{\mathrm{ris}}$ reflection elements can
be expressed as \cite{tang2019wireless}
\begin{equation}
H(r,t)=\sum_{l=1}^{N_{\mathrm{ris}}}\frac{\lambda^{2}}{16\pi^{2}d1_{r,l}d2_{l,t}}e^{-j\left(\frac{2\pi(d1_{r,l}+d2_{l,t})}{\lambda}-\phi_{l}\right)},\label{eq:H_indir}
\end{equation}
where $d1_{r,l}$ is the distance between the $l$-th RIS reflection
element and the $r$-th receive antenna, $d2_{l,t}$ is the distance
between the $t$-th transmit antenna and the $l$-th RIS reflection
element, and $\phi_{l}$ is the phase shift induced by the $l$-th
RIS reflection element}\footnote{\textcolor{black}{Since the assumed inter-antenna separations are
$\lambda/2$, the lengths of the antenna arrays are only a few centimeters.
At the same time, the distances between a RIS and the transmit/receive
antenna array are a few meters (i.e., significantly larger than the
antenna array lengths). Therefore, all transmit signals that originate
from different transmit antennas have almost the same incidence angles
at an arbitrary RIS reflection element. Based on this and the small
length of the receive antenna array, it is reasonable to assume that
we can adjust the reflection angles of each RIS element, so that all
transmit signals reflected from that element can be received by all
receive antennas. In that case, the expression \prettyref{eq:H_indir}
holds true.}}\textcolor{black}{.}

\textcolor{black}{For ease of analysis, we normalize the channel coefficients
by the \ac{FSPL} corresponding to signal transmission from a transmit
antenna placed at height $h_{t,m}$ to a receive antenna placed at
height $h_{r,m}$ via a reflective element at the center of the RIS.
This normalizing \ac{FSPL} is equal to $\lambda^{2}/16\pi^{2}d1_{c}d2_{c}$,
where $d1_{c}=\sqrt{h_{r,m}^{2}+(D-d_{\mathrm{ris}})^{2}}$ and $d2_{c}=\sqrt{h_{t,m}^{2}+d_{\mathrm{ris}}^{2}}$.
Taking into account that $(\forall k,l,n)\;d1_{r,l}d2_{l,t}\approx d1_{1,1}d2_{1,1}$,
the normalized channel coefficients are given as
\begin{equation}
H(r,t)\approx k\sum_{n=1}^{N_{\mathrm{ris}}}\exp\left(-j\frac{2\pi(d1_{r,l}+d2_{l,t})}{\lambda}+j\phi_{l}\right),
\end{equation}
where $k=d1_{c}d2_{c}/d1_{1,1}d2_{1,1}$. With this approximation
in place, the channel matrix $\mathbf{H}$ can be expressed as
\begin{equation}
\mathbf{H}=k\mathbf{V}\mathbf{F}\mathbf{U},\label{eq:INDIR_prod}
\end{equation}
}where we define matrices $\mathbf{U}=\left[\exp(-j2\pi d2_{l,t}/\lambda)\right]\in\mathbb{C}^{N_{\mathrm{ris}}\times N_{t}}$,
$\mathbf{V}=\left[\exp(-j2\pi d1_{r,l}/\lambda)\right]\in\mathbb{C}^{N_{r}\times N_{\mathrm{ris}}}$
and $\mathbf{F}=\mathrm{diag}(\exp(j\phi_{1})\;\cdots\;(\exp(j\phi_{N_{\mathrm{ris}}}))\in\mathbb{C}^{N_{\mathrm{ris}}\times N_{\mathrm{ris}}}$.

\section{Proposed Optimization Schemes \label{sec:Single-stream-trans}}

The channel capacity of single-stream systems is primarily determined
by the receive \ac{SNR} \cite{1284943}. To be able to further increase
the receive \ac{SNR} while keeping the receiver hardware implementation
of the proposed optimization schemes as practical as possible, we
assume a receive signal combining which is realized as a summation
of the signals from all receive antennas, i.e., 
\begin{equation}
y_{\mathrm{sum}}=\sum_{r=1}^{N_{r}}y_{r}=\frac{1}{\sqrt{N_{t}}}\left(\sum_{r=1}^{N_{r}}\sum_{t=1}^{N_{t}}H(r,t)p_{t}\right)s+\sum_{r=1}^{N_{r}}n_{r}.\label{eq:simp_ss_equ}
\end{equation}

In the sequel, we derive two optimization schemes which aim to enhance
the channel capacity of single-stream transmission in indoor mmWave
communications.  The first optimization scheme utilizes only the
phase adjusting capabilities of the \ac{RIS} elements to optimize
the channel capacity. The second optimization scheme jointly uses
the RIS element phase adjusting capabilities \emph{and} the transmit
phase precoding for the optimization of the channel capacity.

\subsection{RIS-only Optimization Scheme\label{subsec:SS-trans-with-prec}}

This optimization scheme is purely based on the RIS element phase
adjusting capabilities, without considering any phase adjustment possibility
of the transmit precoder. Hence, the precoding vector $\mathbf{p}$
is modeled as an all-ones column vector and the resulting receive
signal from \eqref{eq:simp_ss_equ} can be expressed as
\begin{align}
y_{\mathrm{sum}}= & \sum_{r=1}^{N_{r}}y_{r}=\frac{1}{\sqrt{N_{t}}}\Bigg[k\sum_{l=1}^{N_{\mathrm{ris}}}\exp(j\phi_{l})\times\nonumber \\
 & \sum_{r=1}^{N_{r}}\sum_{t=1}^{N_{t}}V(r,l)U(l,t)\Bigg]s+\sum_{r=1}^{N_{r}}n_{r}.\label{eq:rec-sig-lis}
\end{align}
The phase shift value of the \emph{l}-th RIS reflection element should
be chosen to satisfy
\begin{equation}
\phi_{l}=-\arg\left\{ \sum_{r=1}^{N_{r}}\sum_{t=1}^{N_{t}}V(r,l)U(l,t)\right\} .\label{eq:ph_lis}
\end{equation}
Substituting \eqref{eq:ph_lis} into \eqref{eq:rec-sig-lis}, we obtain
\begin{align}
y_{\mathrm{sum}} & =\left(k\sum_{l=1}^{N_{\mathrm{ris}}}\left|\sum_{r=1}^{N_{r}}\sum_{t=1}^{N_{t}}V(r,l)U(l,t)\right|\right)\frac{s}{\sqrt{N_{t}}}+n'\nonumber \\
 & =\frac{1}{\sqrt{N_{t}}}Bs+n',\label{eq:ss-equ}
\end{align}
where the resulting noise signal $n'$ is distributed according to
$\mathcal{CN}(0,N_{0}N_{r})$.

Finally, the channel capacity of this RIS optimization scheme in an
indoor mmWave environment can be calculated as
\begin{equation}
C=\log_{2}\left(1+\frac{B^{2}}{N_{t}N_{r}}\frac{E_{s}}{N_{0}}\right).\label{eq:cap-ris}
\end{equation}

\subsubsection{Approximate Expression}

In this subsection, we derive a closed-form approximate expression\footnote{This expression is valid only when the separations between the adjacent
antennas  are significantly smaller than the distance between the
RIS and the transmit/receive antenna array.} to simplify computation of the channel capacity in \eqref{eq:cap-ris}.
To achieve this, we find an approximate expression for the value $|\sum_{r=1}^{N_{r}}\sum_{t=1}^{N_{t}}V(r,l)U(l,t)|$
which appears in \prettyref{eq:ss-equ}. The following derivation
considers signal transmission via an arbitrary RIS element. Simulation
results in \prettyref{sec:Simulation-Results} show that this approximation
enables very accurate results.

The $l$-th row of $\mathbf{U}$ corresponds to the approximate
response\footnote{In the following channel response expressions, we neglect the influence
of \ac{FSPL}.} of the channel between the transmit antenna array and the $l$-th
RIS reflection element, which is given as \cite[pp. 347--349]{Tse2005}
\begin{align*}
\mathbf{u}_{l,*}\approx & e^{-j\frac{2\pi}{\lambda}(d_{t,l}+\frac{(N_{t}-1)}{2}s_{t}\cos\theta_{t})}\times\\
 & \left[1\;e^{j\frac{2\pi}{\lambda}s_{t}\cos\theta_{t}}\;\cdots\;e^{j\frac{2\pi}{\lambda}(N_{t}-1)s_{t}\cos\theta_{t}}\right]
\end{align*}
where $d_{t,l}$ is the distance between the midpoint of the transmit
antenna array and the $l$-th RIS reflection element, and $\theta_{t}$
is the angle of signal departure at the transmit antenna array (at
the transmit antenna array midpoint). In a similar manner, the $l$-th
column of $\mathbf{V}$ corresponds to the approximate response of
the channel between the $l$-th RIS reflection element and the receive
antenna array, which is given as \cite[pp. 347--349]{Tse2005}
\begin{align*}
\mathbf{v}_{*,l}\approx & e^{-j\frac{2\pi}{\lambda}(d_{r,l}+\frac{(N_{r}-1)}{2}s_{r}\cos\theta_{r})}\times\\
 & \left[1\;e^{j\frac{2\pi}{\lambda}s_{r}\cos\theta_{r}}\;\cdots\;e^{j\frac{2\pi}{\lambda}(N_{r}-1)s_{r}\cos\theta_{r}}\right]^{\mathrm{T}}
\end{align*}
where $d_{r,l}$ is the distance between the $l$-th RIS reflection
element and the midpoint of the receive antenna array, and $\theta_{r}$
is the angle of signal arrival at the receive antenna array (at the
receive antenna array midpoint). Now the channel matrix modeling communication
via the \emph{l-}th RIS reflection element (without optimizing the
phase of the impinging radio waves) can be modeled by
\[
\mathbf{T}^{l}=\mathbf{v}_{*,l}\mathbf{u}_{l,*}.
\]

If we sum all matrix elements from the previous expression, we obtain
\begin{multline*}
\sum_{r=1}^{N_{r}}\sum_{t=1}^{N_{t}}V(r,l)U(l,t)=\sum_{r=1}^{N_{r}}\sum_{t=1}^{N_{t}}T^{l}(r,t)\\
\approx e^{-j\frac{2\pi}{\lambda}(d_{t,l}+\frac{(N_{t}-1)}{2}s_{t}\cos\theta_{t})}e^{-j\frac{2\pi}{\lambda}(d_{r,l}+\frac{(N_{r}-1)}{2}s_{r}\cos\theta_{r})}\\
\times\sum_{m=0}^{N_{t}-1}e^{j\frac{2\pi m}{\lambda}s_{t}\cos\theta_{t}}\sum_{n=0}^{N_{r}-1}e^{j\frac{2\pi n}{\lambda}s_{r}\cos\theta_{r}}\\
=e^{-j\frac{2\pi}{\lambda}(d_{t,l}+d_{r,l})}\frac{\sin\left(\frac{N_{t}\pi}{\lambda}s_{t}\cos\theta_{t}\right)}{\sin\left(\frac{\pi}{\lambda}s_{t}\cos\theta_{t}\right)}\frac{\sin\left(\frac{N_{r}\pi}{\lambda}s_{r}\cos\theta_{r}\right)}{\sin\left(\frac{\pi}{\lambda}s_{r}\cos\theta_{r}\right)}.
\end{multline*}
Finally, an approximate expression that can be used to simplify the
computation of \eqref{eq:cap-ris} is given by
\begin{multline}
\left|\sum_{r=1}^{N_{r}}\sum_{t=1}^{N_{t}}V(r,l)U(l,t)\right|=\\
\left|\frac{\sin\left(\frac{N_{t}\pi}{\lambda}s_{t}\cos\theta_{t}\right)}{\sin\left(\frac{\pi}{\lambda}s_{t}\cos\theta_{t}\right)}\right|\left|\frac{\sin\left(\frac{N_{r}\pi}{\lambda}s_{r}\cos\theta_{r}\right)}{\sin\left(\frac{\pi}{\lambda}s_{r}\cos\theta_{r}\right)}\right|.\label{eq:apr_exp}
\end{multline}

\begin{figure}[t]
\centering{}\includegraphics[width=7cm]{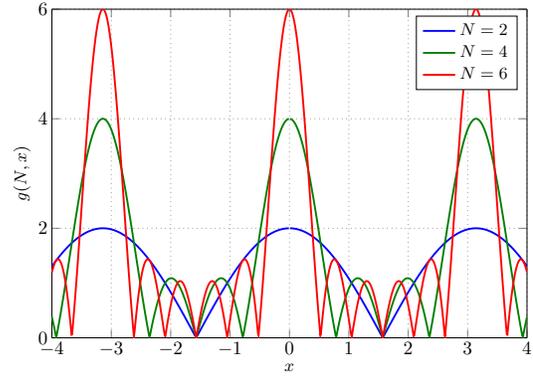}\caption{Auxiliary function $g(N,x)$ versus $N$. \label{fig:Aux_vs_N}}
\end{figure}
Observing the expression \eqref{eq:apr_exp}, we note that its right-hand
side is a product of two terms, each of which can be represented by
an instance of the auxiliary function $g(N,x)=\left|\sin\left(Nx\right)/\sin\left(x\right)\right|$,
which is shown in \prettyref{fig:Aux_vs_N}. Therefore, in the following
we elucidate the key properties of this function in order to provide
insight regarding the behavior of \eqref{eq:apr_exp}. The auxiliary
function $g(N,x)$ has large main lobes that are centered at $x\in\{0,\pm\pi,\pm2\pi,\dots$\},
where the maximum value of this function is achieved. Between the
main lobes, $g(N,x)$ has a number of significantly smaller side lobes.
Different lobes meet at points where the value of $g(N,x)$ is approximately
zero. With increasing $N$, the number of side lobes increases and
the main lobes become narrower. Hence, for a large $N$ there is a
high probability that the value of $g(N,x)$ for some arbitrary $x$
is small. In our particular case, this means that the expression \eqref{eq:apr_exp}
is more likely to have a small value, if $N_{t}$ and/or $N_{r}$
are large. Based on this, we conclude that the RIS-only optimization
scheme is primarily suitable for implementation in communication systems
with a limited number of transmit and receive antennas.

\subsection{Joint Optimization Scheme\label{subsec:Joint ph optLIS-and-Transmit}}

In this subsection, we derive a joint optimization scheme that utilizes
the phase adjusting capabilities of both the RIS elements \emph{and}
the transmit phase precoder. As we aim to have a precoder that is
convenient for hardware implementation for \ac{mmWave} communications,
we consider only phase precoding solutions. Although phase precoding
may not enable the system to achieve the best possible performance,
its low-complexity hardware architecture makes it advantageous in
\ac{mmWave} communications. The transmit phase precoder is modeled
by  the precoding vector  $\mathbf{p}=[\exp(j\beta_{1})\;\cdots\;\exp(j\beta_{N_{t}})]^{\mathrm{T}}$
whose elements are determined by the phase values $\beta_{1},\dots,\beta_{N_{t}}$. 

Since the channel capacity of the joint optimization scheme is given
by
\begin{equation}
C_{\mathrm{joint}}=\log_{2}\left(1+\frac{\left|\sum_{r=1}^{N_{r}}\sum_{t=1}^{N_{t}}H(r,t)\exp(j\beta_{t})\right|^{2}}{N_{t}N_{r}}\frac{E_{s}}{N_{0}}\right),\label{eq:cap-joint}
\end{equation}
our optimization goal can be expressed as
\begin{equation}
\mathop{\mathop{\text{max}}}_{\mathbf{p},\mathbf{F}}\;\left|\sum_{r=1}^{N_{r}}\sum_{t=1}^{N_{t}}H(r,t)\exp(j\beta_{t})\right|.\label{eq:ss_opt}
\end{equation}
First, we choose the transmit precoder phase values $\beta_{1},\dots,\beta_{N_{t}}$
so that all terms $\sum_{r=1}^{N_{r}}H(r,t)\exp(j\beta_{t})$ are
co-phased; thus we obtain\textcolor{black}{
\begin{equation}
\beta_{t}=-\arg\left\{ \sum_{r=1}^{N_{r}}H(r,t)\right\} .
\end{equation}
Now the optimization problem becomes 
\begin{multline}
[\hat{\phi}_{1},\dots,\hat{\phi}_{N_{\mathrm{ris}}}]=\mathop{\mathop{\text{arg max}}}_{\substack{\phi_{1},\dots,\phi_{N_{\mathrm{ris}}}\in[0,2\pi]}
}\sum_{t=1}^{N_{t}}\left|\sum_{r=1}^{N_{r}}H(r,t)\right|,\label{eq:opt_prob_init}
\end{multline}
and it can be rewritten as
\begin{align}
[\hat{\phi}_{1},\dots,\hat{\phi}_{N_{\mathrm{ris}}}] & =\mathop{\mathop{\text{arg max}}}_{\substack{\phi_{1},\dots,\phi_{N_{\mathrm{ris}}}\in[0,2\pi]}
}\sum_{t=1}^{N_{t}}\left|\mathbf{1}\mathbf{V}\mathbf{F}\mathbf{u}_{*,t}\right|,\label{eq:lis_opt_prob}
\end{align}
}where $\mathbf{1}$ is an $1\times N_{r}$ all-ones vector. We note
that the optimization problem \eqref{eq:lis_opt_prob} is analytically
intractable and any exhaustive search solution incurs an extremely
large search space. To overcome this issue, we introduce a simple
sub-optimal technique to solve this optimization problem. Each summation
term in \eqref{eq:lis_opt_prob} can be expressed as
\begin{equation}
\left|\mathbf{1}\mathbf{V}\mathbf{F}\mathbf{u}_{*,t}\right|=\left|\sum_{l=1}^{N_{\mathrm{ris}}}\left(\sum_{r=1}^{N_{r}}V(r,l)\right)U(l,t)e^{j\phi_{l}}\right|.\label{eq:sing_opt_term}
\end{equation}
We observe that the expression \eqref{eq:sing_opt_term} achieves
its maximum when the RIS element phase shift values $\phi_{1},\dots,\phi_{N_{\mathrm{ris}}}$
enable co-phasing of $N_{\mathrm{ris}}$ individual summation terms.
Since the optimization problem \eqref{eq:lis_opt_prob} consists a
sum of $N_{t}$ different expressions \eqref{eq:sing_opt_term}, it
is not possible to choose $\phi_{1},\dots,\phi_{N_{\mathrm{ris}}}$
so that we have co-phasing in all $N_{t}$ expressions. Instead, we
define the technique of \emph{global co-phasing} which we implement
for each $\phi_{l}$ independently. By \emph{global co-phasing} we
imply adjusting $\phi_{l}$ so that the sum of the phase deviations
of the $N_{t}$ terms influenced by $\phi_{l}$ in \eqref{eq:lis_opt_prob}
from the target phase has the minimum absolute value. For convenience
we take that the target phase is 0. If we define
\[
\delta_{t,l}=\arg\left\{ \left(\sum_{r=1}^{N_{r}}V(r,l)\right)U(l,t)\right\} ,
\]
then the sum of the phase deviations is given by $\sum_{t=1}^{N_{t}}(\delta_{t,l}+\phi_{l})=\sum_{t=1}^{N_{t}}\delta_{t,l}+N_{t}\phi_{l}.$
Since our goal is that the previous expression is 0, we finally obtain
\begin{equation}
\phi_{l}=-\frac{1}{N_{t}}\sum_{t=1}^{N_{t}}\delta_{t,l}.
\end{equation}

It should be noted that the amplitude of the term influenced by $\phi_{l}$
in \eqref{eq:lis_opt_prob} is determined by $\sum_{r=1}^{N_{r}}V(r,l)$.
Also, this sum can be expressed in the form $g(N,x)$, where $N=N_{r}$.
Therefore, the amplitude of the term influenced by $\phi_{l}$ in
\eqref{eq:lis_opt_prob} is more likely to be small for larger $N_{r}$
and only very limited capacity gains can be generally achieved when
global co-phasing is implemented in this case. For this reason, the
considered phase optimization scheme is most suitable for communication
systems with a limited number of receive antennas.

\section{Simulation Results \label{sec:Simulation-Results}}

\begin{figure*}[t]
\begin{centering}
\subfloat[$N_{t}=8$, $N_{r}=4$ and $N_{\mathrm{ris}}=50$.]{\centering{}\includegraphics{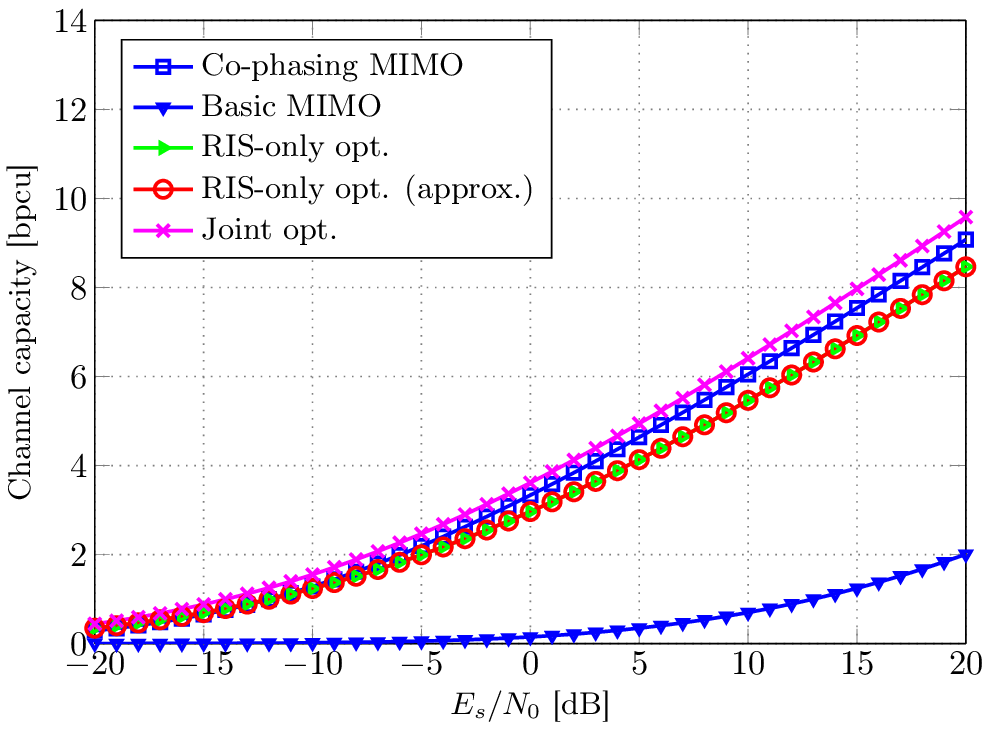}}\subfloat[$N_{t}=8$, $N_{r}=2$ and $N_{\mathrm{ris}}=50$. \label{fig:glob-co-vs-RIS}]{\centering{}\includegraphics{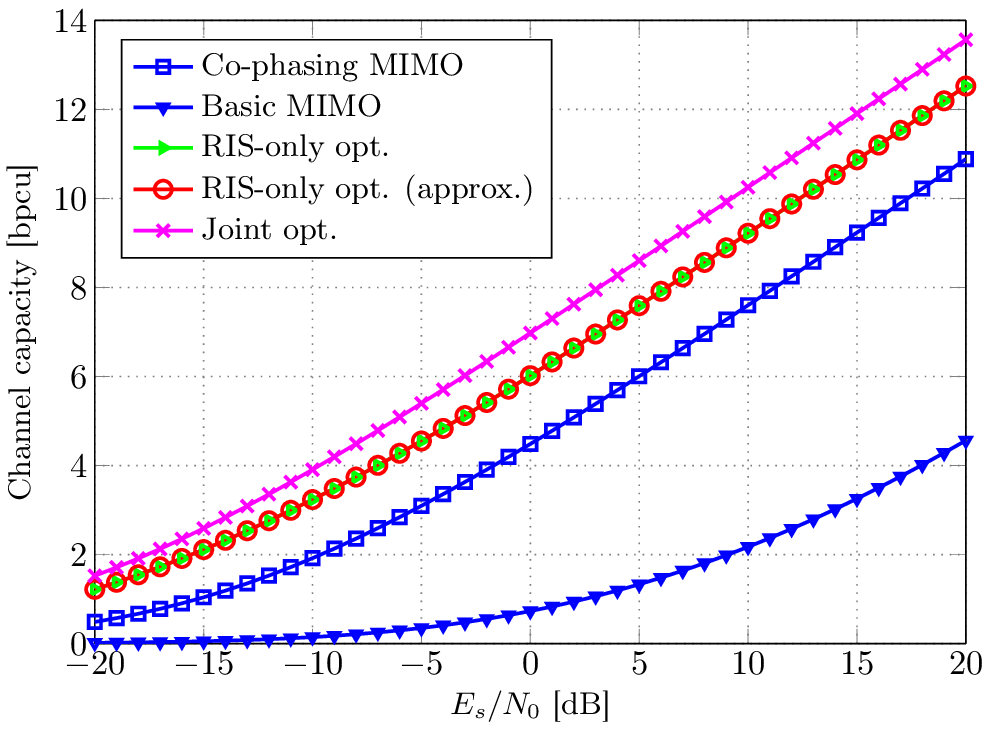}}
\par\end{centering}
\begin{centering}
\subfloat[$N_{t}=8$, $N_{r}=4$ and $N_{\mathrm{ris}}=100$. \label{fig:max_ris_elem}]{\centering{}\includegraphics{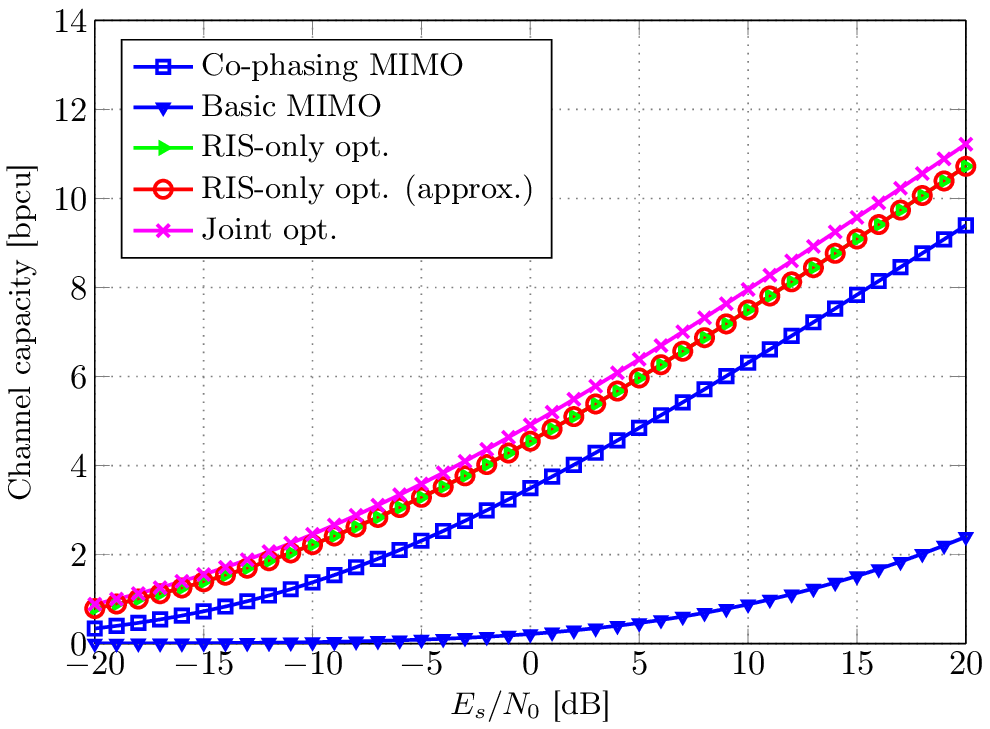}}\subfloat[$N_{t}=16$, $N_{r}=4$ and $N_{\mathrm{ris}}=100$. \label{fig:glob-co-vs-RIS-1}]{\centering{}\includegraphics{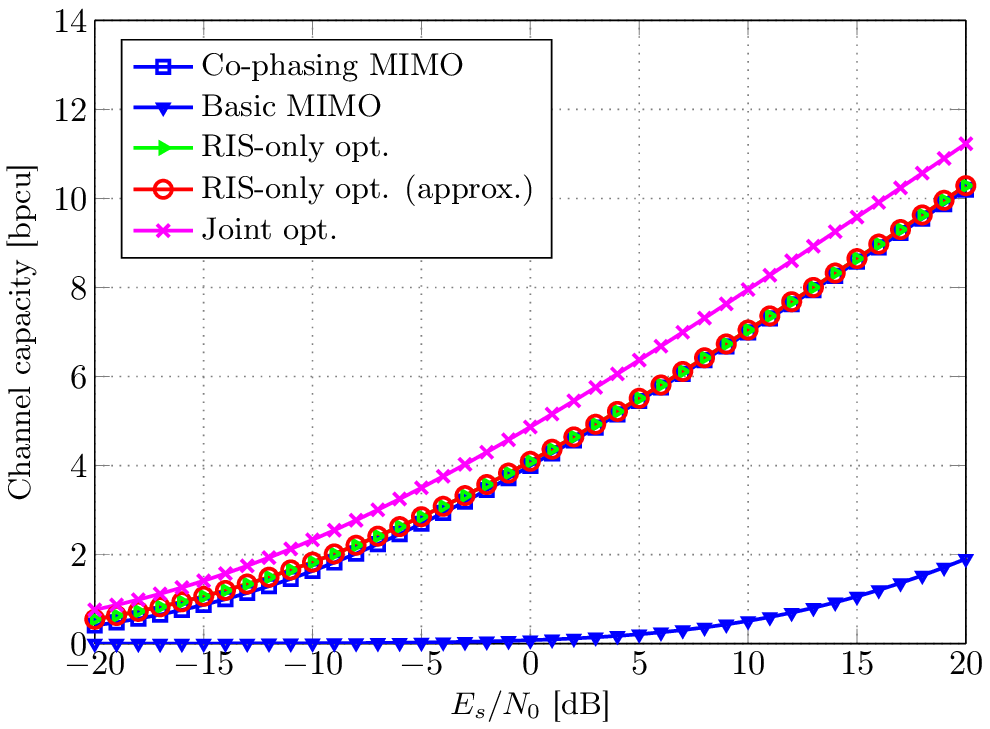}}
\par\end{centering}
\centering{}\caption{Channel capacity of the RIS-only optimization scheme \prettyref{eq:cap-ris}
and the joint optimization scheme \prettyref{eq:cap-joint}, versus
the benchmark schemes (i.e., co-phasing MIMO \prettyref{eq:cap co-phasing}
and basic MIMO \prettyref{eq:cap-basic}).\label{fig:Cap-comp}}
\end{figure*}
In this section, we present channel capacity simulation results for
the RIS-only optimization scheme and the joint optimization scheme.
We assume an indoor \ac{mmWave} environment where no \ac{LOS} path
exists, and where the only \ac{NLOS} paths are those provided by
the RIS\footnote{Equivalently, other NLOS paths are assumed to provide negligible gain
relative to that provided by the RIS.}. In order to quantify the capacity gain due to the RIS optimization,
we consider two benchmark schemes which utilize an RIS for reflecting
the transmitted signal towards the receive antennas, but do not exploit
the phase adjustment capabilities of the RIS. The first benchmark
scheme, called \emph{co-phasing} MIMO, is based on the implementation
of only phase precoding at the transmitter and the receiver. The appropriate
precoder phase adjustments and the channel capacity expression for
this scheme are provided in \prettyref{app:App-co-phasing}. The second
benchmark scheme, called \emph{basic} MIMO, has the simplest architecture
among the considered schemes, and it does not use transmit/receive
precoding or \textcolor{black}{RIS phase adjustment }to optimize the
channel capacity. The channel capacity of this scheme can be calculated
via
\begin{equation}
C_{\mathrm{basic}}=\log_{2}\left[1+\frac{\left|\sum_{r=1}^{N_{r}}\sum_{t=1}^{N_{t}}H(r,t)\right|^{2}}{N_{t}N_{r}}\frac{E_{s}}{N_{0}}\right].\label{eq:cap-basic}
\end{equation}

In the simulations, the setup parameters are $\lambda=5\,\mathrm{mm}$
(i.e., $f=60\,\mathrm{GHz}$), $s_{t}=s_{r}=\lambda/2=2.5\,\mathrm{mm}$,
$s_{\mathrm{ris}}=\lambda/2=2.5\,\mathrm{mm}$, $D=5\,\mathrm{m}$
and $d_{\mathrm{ris}}=2.5\,\mathrm{m}$. To obtain channel capacity
results that are independent of a specific communication system geometry,
we vary the antenna array heights $h_{t}$ and $h_{r}$. More precisely,
$h_{t}$ is chosen from a uniform distribution between 2\,m to 3\,m
with a resolution of 2\,cm (i.e., $h_{t,m}=2.5\,\mathrm{m}$), and
$h_{r}$ is chosen from a uniform distribution between 0.8\,m to
1.8\,m with a resolution of 2\,cm (i.e., $h_{r,m}=1.3\,\mathrm{m}$).
The channel capacity is then averaged using a Monte Carlo approach.

\textcolor{black}{In \prettyref{fig:Cap-comp}, we show the channel
capacity simulation results for the RIS-only optimization scheme and
the joint optimization scheme, versus the benchmark schemes (i.e.,
co-phasing MIMO and basic MIMO). In all cases, the joint optimization
scheme achieves a higher channel capacity than the other considered
schemes. As expected, the RIS-only optimization scheme has a lower
channel capacity than the joint optimization scheme because it uses
only the RIS elements for phase adjustment. However, with the increase
of the number of RIS elements the channel capacity of the RIS-only
optimization scheme becomes almost equal to the channel capacity of
the joint optimization scheme (see \prettyref{fig:max_ris_elem}).
In that case, the number of RIS elements is much larger than the number
of transmit antennas, so that any channel capacity gain obtained through
the transmit precoding is negligibly small compared to the channel
capacity gain obtained through phase adjustment of the RIS elements.
Increasing the number of receive antennas causes a channel capacity
reduction for both optimization schemes, due to the reasons already
explained in \prettyref{sec:Single-stream-trans}. The same is valid
for the number of transmit antennas of the RIS-only optimization scheme.
Also, we notice that the approximate expression \prettyref{eq:apr_exp}
for the channel capacity of the RIS-only optimization scheme gives
approximately the same results as the exact channel capacity expression.
Among the two benchmark schemes, basic MIMO always achieves the lower
channel capacity because of the lack of hardware capabilities to optimize
the channel capacity. \balance}

\section{\textcolor{black}{Conclusion}}

\textcolor{black}{In this paper, we studied the channel capacity optimization
of single-stream transmission utilizing \acp{RIS} in mmWave indoor
environments without any LOS path, and proposed two optimization schemes.
For the first optimiz}ation scheme, which targets only phase adjustments
at the RIS, we derived an approximate expression which simplifies
the channel capacity calculation and explains the connection between
the achievable channel capacity gains and the system parameters. \textcolor{black}{For
the second optimization scheme that jointly utilizes the phase adjusting
capabilities of both the RIS elements and the transmit phase precoder,
we developed a low-complexity technique to obtain the phase shift
values of the \ac{RIS} elements. Simulation results show that both
optimization schemes produce a very significant channel capacity gain,
and that this gain increases with the number of RIS elements.}

\appendices{}

\section{Phase Adjustments for Co-phasing MIMO\label{app:App-co-phasing}}

Since co-phasing MIMO is based on the use of transmit and receive
phase precoding, the resulting receive signal after the receive signal
combining is given as
\begin{equation}
y_{\mathrm{sum}}=\frac{1}{\sqrt{N_{t}}}\mathbf{r}^{\mathrm{T}}\mathbf{H}\mathbf{t}s+\sum_{r=1}^{N_{r}}n_{r},\label{eq:cophasing-sys-mod}
\end{equation}
where $\mathbf{r}=[e^{j\alpha_{1}}\;\cdots\;e^{j\alpha_{N_{r}}}]^{\mathrm{T}}$
and $\mathbf{t}=[e^{j\gamma_{1}}\;\cdots\;e^{j\gamma_{N_{t}}}]^{\mathrm{T}}$.
Based on \prettyref{eq:cophasing-sys-mod}, the channel capacity expression
for co-phasing MIMO can be written as
\begin{equation}
C_{\mathrm{co-phasing}}=\log_{2}\left(1+\frac{\left|\mathbf{r}^{\mathrm{T}}\mathbf{H}\mathbf{t}\right|^{2}}{N_{t}N_{r}}\frac{E_{s}}{N_{0}}\right).\label{eq:cap co-phasing}
\end{equation}

As the channel capacity is primarily determined by the value of $\left|\mathbf{r}^{\mathrm{T}}\mathbf{H}\mathbf{t}\right|$,
the considered phase precoders should be designed to maximize that
value. First, we choose $\alpha_{1},\dots,\alpha_{N_{r}}$ to co-phase
the receive antenna signals before their combining (addition) and
thus we obtain
\begin{equation}
\alpha_{r}=-\arg\left\{ \mathbf{h}_{r,*}\mathbf{t}\right\} .
\end{equation}
Now we have
\[
\mathbf{r}^{\mathrm{T}}\mathbf{H}\mathbf{t}=\sum_{r=1}^{N_{r}}\left|\mathbf{h}_{r,*}\mathbf{t}\right|.
\]
Applying the technique of\emph{ }global co-phasing, which is previously
defined in Subsection \prettyref{subsec:Joint ph optLIS-and-Transmit},
to the elements of $\mathbf{t}$ we finally obtain
\begin{equation}
\gamma_{t}=-\frac{1}{N_{t}}\sum_{r=1}^{N_{r}}\arg\left\{ H(r,t)\right\} .
\end{equation}

\section*{}

\bibliographystyle{IEEEtran}
\bibliography{IEEEabrv,IEEEexample,RIS_indoor_fin}

\begin{IEEEbiography}[{\fbox{\begin{minipage}[t][1.25in]{1in}%
Replace this box by an image with a width of 1\,in and a height of
1.25\,in!%
\end{minipage}}}]{Your Name}
 All about you and the what your interests are.
\end{IEEEbiography}

\begin{IEEEbiographynophoto}{Coauthor}
Same again for the co-author, but without photo
\end{IEEEbiographynophoto}

\end{document}